\documentclass[11pt]{article}

\usepackage[margin=1in]{geometry} 
\usepackage{cancel}
\usepackage{array}
\newcolumntype{L}[1]{>{\raggedright\arraybackslash}p{#1}}
\usepackage{graphicx}     
\usepackage{xcolor}       
\usepackage{float}
\usepackage{booktabs}
\usepackage{enumerate}
\usepackage{centernot}
\usepackage{setspace}
\usepackage{fancyhdr}
\usepackage{subcaption}
\usepackage{abstract}
\usepackage[skip=2pt]{caption}

\usepackage{amsfonts, amsthm, amsmath, amssymb, gensymb}
\usepackage{newtxtext}
\usepackage{newtxmath}
\usepackage{stmaryrd}
\usepackage{mathrsfs}
\usepackage{mathtools}
\usepackage[bb=boondox]{mathalfa}
\everymath{\displaystyle}
\allowdisplaybreaks

\providecommand{\keywords}[1]{%
  \par\smallskip
  \noindent\textbf{Keywords:} #1\par
}

\usepackage[backend=biber,style=numeric]{biblatex}
\usepackage[colorlinks=true, linkcolor=blue, citecolor=blue, urlcolor=blue]{hyperref}

\definecolor{PrimaryBlue}{RGB}{29,79,145}        
\definecolor{SecondaryBlue1}{RGB}{0,48,135}      
\definecolor{SecondaryBlue2}{RGB}{0,119,200}     
\definecolor{SecondaryBlue3}{RGB}{108,172,228}   

\definecolor{Neutral1}{RGB}{34,34,34}       
\definecolor{Neutral2}{RGB}{83,86,90}       
\definecolor{Neutral3}{RGB}{208,208,206}    
\definecolor{Neutral4}{RGB}{240,240,240}    

\definecolor{Accent1}{RGB}{199,80,0}   

\definecolor{Transactional1}{RGB}{142,75,53}   
\definecolor{Transactional2}{RGB}{220,42,42}   

\definecolor{DodgerBlue}{RGB}{0,162,255}
\title{From Headlines to Holdings: \\ Deep Learning for Smarter Portfolio Decisions}

\author{
  \begin{tabular}{ccc}
    Yun Lin$^{*}$ & Jiawei Lou$^{*}$ & Jinghe Zhang$^{*}$ \\
    {\small \texttt{yl5852@barnard.edu}} & 
    {\small \texttt{jl6685@barnard.edu}} & 
    {\small \texttt{jz3893@columbia.edu}}
  \end{tabular}
}
\date{July 2025}

\begin{document}
\onehalfspacing
\maketitle
\renewcommand{\thefootnote}{}\footnotetext{$^{*}$Authors are listed in alphabetical order and contributed equally to this work.
The authors thank Professor George Dragomir, Professor Dobrin Marchev, and Vihan Pandey for continuous guidance and feedback during every stage of our research. We are grateful for the Columbia University Department of Mathematics, the CSUREMM Program, and Barnard SRI for funding our project.}
\renewcommand{\thefootnote}{\arabic{footnote}}
\begin{abstract}
    Deep learning offers new tools for portfolio optimization. We present an end-to-end framework that directly learns portfolio weights by combining Long Short-Term Memory (LSTM) networks to model temporal patterns, Graph Attention Networks (GAT) to capture evolving inter-stock relationships, and sentiment analysis of financial news to reflect market psychology. Unlike prior approaches, our model unifies these elements in a single pipeline that produces daily allocations. It avoids the traditional two-step process of forecasting asset returns and then applying mean–variance optimization (MVO), a sequence that can introduce instability. We evaluate the framework on nine U.S. stocks spanning six sectors, chosen to balance sector diversity and news coverage. In this setting, the model delivers higher cumulative returns and Sharpe ratios than equal-weighted and CAPM-based MVO benchmarks. Although the stock universe is limited, the results underscore the value of integrating price, relational, and sentiment signals for portfolio management and suggest promising directions for scaling the approach to larger, more diverse asset sets.
\end{abstract}

\keywords{Portfolio Optimization, Asset Allocation, Graph Neural Networks, Graph Attention Networks, Long Short-Term Memory, Financial News Sentiment}
\newpage

\begingroup
\setstretch{0.88}

\tableofcontents
\endgroup

\newpage
\section{Introduction and Motivation} 
Portfolio optimization has long been a central problem in finance, where the objective is to allocate capital across assets in a way that balances risk and return. Traditional approaches usually follow a two-step process~\cite{zhou2023two_stage_portfolio}: first, forecasting the returns or prices of individual assets, and second, applying Mean-Variance Optimization (MVO) to determine portfolio weights. Although widely used, this framework faces two important limitations. It relies heavily on the accuracy of return forecasts, meaning that any prediction errors can be amplified in the optimization stage. Moreover, it treats assets independently, overlooking the interdependencies that are fundamental to real financial markets.

Recent advances in deep learning provide new opportunities to overcome these challenges. Long Short-Term Memory (LSTM) networks are well-suited to capturing temporal dependencies in financial time series~\cite{chalvatzis2019high}, while Graph Neural Networks (GNNs), and particularly Graph Attention Networks (GATs), can model the relationships among assets~\cite{ekmekcioglu2023gnnportfolio, pacreau2021gnn}. Several recent studies have shown the promise of these techniques for portfolio allocation. Lu et al.~\cite{lu2025leveraging} introduced a multilayer LSTM-GAT-AM model with dual graph structures that outperformed benchmark strategies. Zhang et al.~\cite{zhang2020deep} proposed an end-to-end LSTM framework that surpassed MVO and proved more stable than two-stage designs such as Lu’s. Korangi et al.~\cite{korangi2024large} further highlighted the advantages of GAT-based approaches for capturing complex inter-asset interactions. Beyond architecture, alternative modeling choices have also been shown to matter: Pacreau et al.~\cite{pacreau2021gnn} emphasized the value of dynamically updated graphs to reflect evolving market conditions, and Srinivas et al.~\cite{srinivas2023effects} demonstrated that sentiment data from financial news can significantly improve predictive accuracy.

Building on these insights, our work develops an end-to-end framework that unifies three elements: temporal modeling with LSTMs, relational modeling with GATs, and sentiment analysis of financial news. Unlike traditional methods that separate prediction from optimization, our model learns portfolio weights directly, reducing the risk of compounded errors. To demonstrate the approach, we focus on a set of nine U.S. stocks across six major sectors: Apple and Nvidia (Information Technology), Johnson \& Johnson and Thermo Fisher Scientific (Health Care), Tesla and Amazon (Consumer Discretionary), Boeing (Industrials), Costco (Consumer Staples), and Valero Energy (Energy). This universe was selected from a larger pool of S\&P~500 companies to ensure sector coverage and diversification while maintaining sufficient news coverage for sentiment analysis.

The objective of this study is twofold: first, to evaluate whether this integrated LSTM-GAT framework can outperform benchmark strategies such as equal-weighted portfolios, CAPM-based MVO, and the conventional two-step pipeline; and second, to assess the contribution of dynamic graph structures and sentiment features to both portfolio performance and stability.

\section{Methodology}
Our framework integrates temporal modeling, relational modeling, and portfolio optimization in a unified pipeline. This section introduces the main components in turn: the Long Short-Term Memory (LSTM) network for temporal dependencies, the Graph Attention Network (GAT) for cross-asset relationships, the construction of static and dynamic graphs, the Sharpe ratio--based objective function, and the overall architecture of the model.

\subsection{Long Short-Term Memory}
Long Short-Term Memory (LSTM) networks, introduced by Hochreiter and Schmidhuber~\cite{hochreiter1997lstm}, are a class of recurrent neural networks designed to model sequential data and capture long-term dependencies. Standard RNNs often fail to learn from long sequences due to vanishing or exploding gradients. LSTMs address this limitation with gating mechanisms, such as \textit{input}, \textit{forget}, and \textit{output} gates, that regulate information flow and preserve relevant signals over time. Because of these properties, LSTMs have been widely adopted in time series forecasting and natural language processing. In our framework, the LSTM captures temporal dynamics in stock-level features such as returns, volatility, and sentiment scores.

\subsection{Graph Attention Network}
Graph Attention Networks (GATs), proposed by Veli\v{c}kovi\'c et al.~\cite{veličković2018gat}, extend graph neural networks by incorporating attention mechanisms into message passing. Each node learns to assign different levels of importance to its neighbors, allowing the network to focus on the most informative relationships. Unlike earlier GNN models, GATs avoid expensive global operations such as eigen decomposition, making them more scalable to dynamic or large graphs. In financial applications, GATs have been shown to capture complex inter-asset dependencies~\cite{lu2025leveraging}, which are often missed by models treating each stock in isolation. In our framework, the GAT refines the LSTM embeddings by incorporating information from related assets.

\subsection{Graph Construction}\label{graph}
To apply GAT, we define graphs that represent the relationships between stocks. The following two types are considered.

\paragraph{Static Graph.} The static graph remains fixed throughout training and evaluation. It is constructed from the correlation matrix of daily log returns over the entire training period, with edge weights given by correlation coefficients ranging from $-1$ to $1$. Correlations of log returns are a standard measure of asset co-movement and capture long-term linear dependencies in price dynamics, allowing the graph structure to reflect persistent interactions between stocks.

\paragraph{Dynamic Graph.} The dynamic graph captures short-term, evolving dependencies. It is updated weekly to balance responsiveness with computational efficiency. A binary edge is drawn between two assets if they (i) belong to the same GICS sector, (ii) exhibit an absolute correlation in 5-day returns above 0.5, or (iii) exhibit an absolute correlation in 5-day sentiment scores above 0.5. This dynamic structure allows the model to adapt to shifting market conditions while preserving sectoral information.

\subsection{Objective Function} \label{loos function}
Unlike traditional pipelines that predict returns and then apply optimization separately, our model is trained end-to-end by directly maximizing portfolio performance. Specifically, we use a Sharpe ratio--based loss function:
\begin{equation}
\mathcal{L}(w, r, \Sigma) \;=\; -\frac{w^{\!\top} r}{\sqrt{\,w^{\!\top} \Sigma\,w }}.
\label{Loss}
\end{equation}
where $w$ is the portfolio weight vector, $r$ the realized return vector for the next period, and $\Sigma$ the covariance matrix of returns.

Optimizing directly for the Sharpe ratio is advantageous because it aligns the training objective with the ultimate performance criterion in portfolio management. This avoids the compounding errors that arise when forecasts of individual returns are optimized separately in a second stage.

\subsection{Model Architecture}
\begin{figure}[H]
  \centering
  \includegraphics[width=\linewidth]{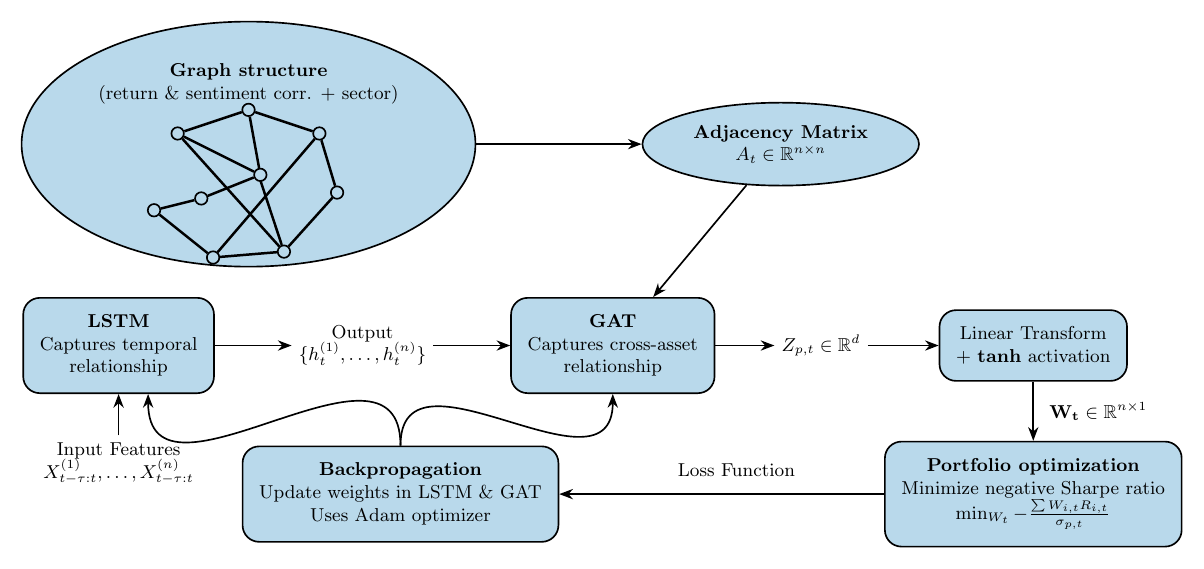}
  \caption{End-to-end LSTM-GAT portfolio optimization framework}
  \label{fig:model_architecture}
\end{figure}

The architecture integrates the components described in Figure~\ref{fig:model_architecture}. 

At each time step $t$, each asset $i$ has a feature matrix $X^{(i)}_{t-r:t}$ corresponding to a lookback window of length $r$. In our experiments, $r$ is set to 30 trading days, reflecting a balance between capturing medium-term dynamics and computational efficiency. Features include both price-based metrics (returns, volatility, momentum indicators) and sentiment-based metrics (average sentiment, variance).

The workflow proceeds as follows:

\begin{enumerate}
    \item \textbf{Temporal encoding.} Each asset’s feature sequence is processed through a shared LSTM, producing hidden states $h^{(i)}_t \in \mathbb{R}^H$.
    \item \textbf{Relational encoding.} The hidden states are passed to a GAT, which aggregates cross-asset information using the static or dynamic graph, generating refined embeddings $z^{(i)}_t \in \mathbb{R}^D$.
    \item \textbf{Weight generation.} Refined embeddings are fed into a linear layer with a $\tanh$ activation to produce raw weights $W_t \in \mathbb{R}^n$.
    \item \textbf{Normalization.} Portfolio weights are scaled as
    \begin{equation}
        w_{i,t} = \frac{W_{i,t}}{\sum_{j=1}^n W_{j,t}},
    \end{equation}
    ensuring that allocations sum to one and allowing both long and short positions.
    \item \textbf{Optimization.} Parameters are updated end-to-end using the Adam optimizer to minimize the negative Sharpe ratio loss.
\end{enumerate}

This architecture enables the model to learn directly from both temporal signals and inter-asset relationships, producing portfolio allocations that are jointly optimized for risk-adjusted returns.

\section{Experiment}

\subsection{Selection of Stocks}
Our experiments focus on a fixed universe of nine U.S. stocks drawn from the S\&P~500. While this set is small relative to professional investment universes, it provides a controlled environment for testing our framework and serves as a proof of concept. Stocks were chosen according to three criteria: (i) low pairwise return correlations to encourage diversification, (ii) representation across multiple sectors, and (iii) broad news coverage to ensure reliable sentiment signals.

From an initial pool of 50 widely followed S\&P~500 companies,  we selected nine stocks spanning six sectors: Apple (AAPL) and Nvidia (NVDA, Information Technology); Johnson \& Johnson (JNJ) and Thermo Fisher Scientific (TMO, Health Care); Tesla (TSLA) and Amazon (AMZN, Consumer Discretionary); Boeing (BA, Industrials); Costco (COST, Consumer Staples); and Valero Energy (VLO, Energy).

\subsection{Data Collection and Cleaning}
Building the future set required collecting both price and news data. Daily price data (open, high, low, close, and volume) were obtained from the \textbf{AlphaVantage API}~\cite{alphavantage} for the period January 2021 to May 2025, yielding approximately 1,080 trading days per stock. News data were retrieved from the \textbf{MarketAux API}~\cite{marketaux_api}, which provides timestamped articles with metadata including title, snippet, relevance score, and sentiment score (ranging from $-1$ for strongly negative to $+1$ for strongly positive). On average, 35--50 articles per stock per month were available during this period.

For each stock and trading day, we aggregated the number of matched articles and computed the average sentiment score. Articles published on non-trading days were shifted to the next trading day.

Benchmark data were also collected: S\&P~500 index levels (via the yfinance API) and 3-month U.S. Treasury yields (via FRED~\cite{capm}) to construct CAPM-MVO baselines. After preprocessing, the datasets were merged into a unified panel with both price- and sentiment-based features (see Appendix \ref{feature} for definitions).

\subsection{Assumptions}
Several simplifying assumptions are made in this study. The investable universe is restricted to the nine selected stocks, with no rebalancing into new assets. Trades are assumed to execute perfectly at official open or close prices, with zero slippage or delay. All trading data are treated as correctly timestamped and free of missing values. Sector membership is assumed constant, ignoring possible corporate restructuring. Finally, transaction costs, bid-ask spreads, and market impacts are excluded.

These assumptions are common in exploratory studies but may result in optimistic outcomes. In real-world settings, transaction costs and liquidity constraints would lower realized returns, and dynamic sector reclassifications could alter graph connectivity. Our results should therefore be interpreted as upper-bound estimates of the framework’s potential.

\subsection{Features and Models}
We evaluate five variants of the LSTM-GAT framework:

\begin{itemize}
\item \textbf{Model v1 (baseline):} close price, volume, log return; static graph
\item \textbf{Model v2:} adds annualized returns, 5-day rolling volatility, MACD
\item \textbf{Model v3:} further adds sentiment-based features (sentiment variance, weighted sentiment)
\item \textbf{Model v4:} same as v3 but with dynamic graphs updated weekly
\item \textbf{Model v5:} applies Principal Component Analysis (PCA) to v4’s features, retaining six principal components out of twelve features.
\end{itemize}

PCA is used in Model v5 to reduce dimensionality and noise, thereby improving model stability during periods of high volatility. Full feature definitions and calculations are provided in Appendix \ref{feature}.

\begin{table}[H]
\centering
\caption{Feature usage and graph type across model versions}
\begin{tabular}{lccccc}
\toprule
\textbf{Feature / Graph} & \textbf{Model v1} & \textbf{Model v2} & \textbf{Model v3} & \textbf{Model v4} & \textbf{Model v5} \\
\midrule
\textbf{Graph Type} & Static & Static & Static & Dynamic & Dynamic \\
1. Close/Volume & \checkmark & \checkmark & \checkmark & \checkmark & \checkmark \\
2. Log Return & \checkmark & \checkmark & \checkmark & \checkmark & \checkmark \\
3. Annualized Returns (1W/2W/1M) &  & \checkmark & \checkmark & \checkmark & \checkmark \\
4. 5D Rolling Volatility &  & \checkmark & \checkmark & \checkmark & \checkmark \\
5. MACD (1W–1M) &  & \checkmark & \checkmark & \checkmark & \checkmark \\
6. News Count &  &  &  &  & \checkmark \\
7. Average Sentiment &  &  &  &  & \checkmark \\
8. Sentiment Variance &  &  & \checkmark & \checkmark & \checkmark \\
9. Weighted Sentiment &  &  & \checkmark & \checkmark & \checkmark \\
\bottomrule
\end{tabular}
\raggedright
\label{tab:model_features}
\end{table}

\subsection{Training, Validation, and Hyperparameter Tuning}

We split the dataset into three parts: 70\% for training and hyperparameter tuning, 20\% of that training portion for internal validation, and the final 30\% for out-of-sample testing. The test period spans early 2024 to mid-2025 and includes the April 2025 tariff-induced market shock, an event that led to heightened volatility and thus provides a meaningful scenario for evaluating model robustness under stress. 
\begin{figure}[H]
    \centering
    \includegraphics[width=1\linewidth]{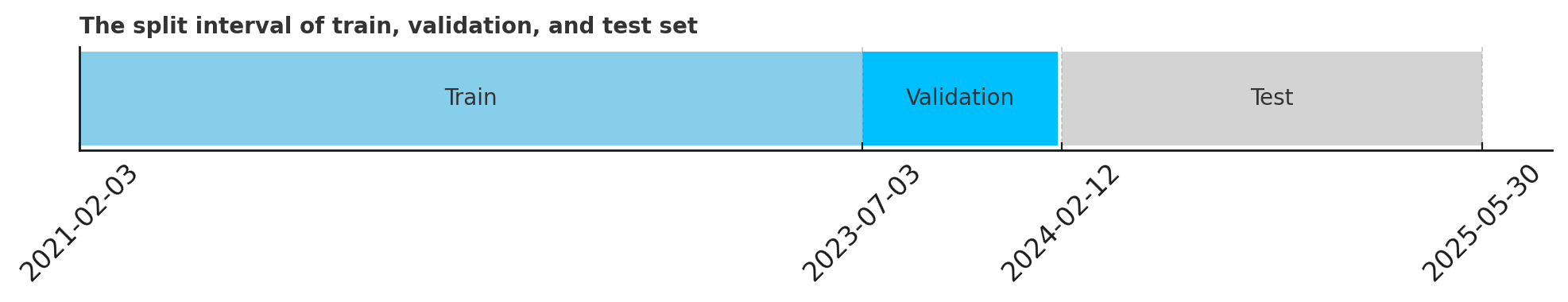}
    \caption{The split interval of train, validation, and test set}
    \label{fig:interval_split}
\end{figure}
During training, we use the Adam optimizer, a widely used method for efficient and stable neural-network training~\cite{kingma2014adam}, to update model weights. Hyperparameter tuning is conducted with \textbf{Optuna} over 50 trials, searching across a predefined space (see Appendix~\ref{tab:hyperparam_ranges} for the full range of hyperparameters).

All input features are standardized per ticker using a pre-fitted StandardScaler. The training process runs for up to 40 epochs, using randomly sampled mini-batches of trading dates. The dynamic graph structure is refreshed every five trading days to capture evolving relationships. At each step, the model predicts portfolio weights, computes the negative Sharpe ratio loss, and updates parameters via backpropagation.

After each epoch, performance is evaluated on the validation set to guide hyperparameter selection, with the Sharpe ratio serving as the primary criterion. Once the best configuration is identified, the model is retrained on the full 70\% training block using the chosen hyperparameters.

Finally, the trained model with optimized hyperparameters is evaluated on the held-out 30\% test set. We report a comprehensive set of metrics: annualized Sharpe ratio, annualized volatility, cumulative and annualized returns, Value at Risk (VaR), and maximum drawdown. For reproducibility, we fix the random seed at 42 across all components and run the experiments on CPU. Results may vary slightly across different hardware.

\section{Results and Discussion}
\subsection{Results and Comparison}

\begin{figure}[H]
    \centering
    \includegraphics[width=0.8\linewidth]{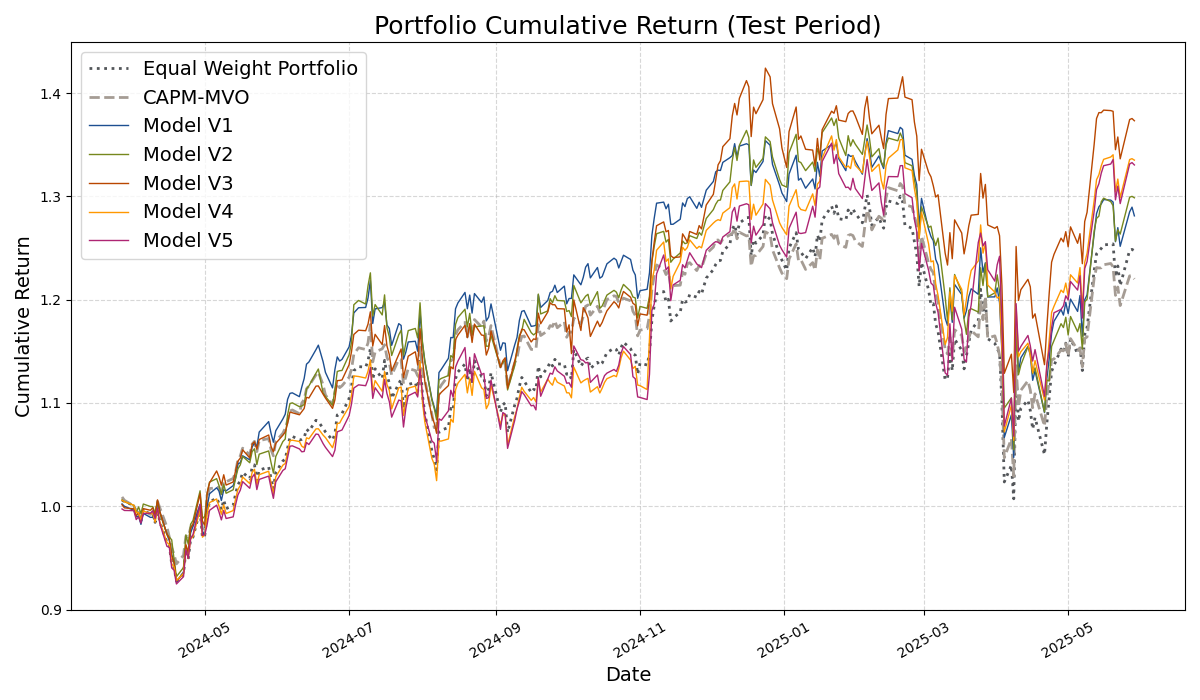}
    \caption{Portfolio cumulative return comparison over the test period.}
    \label{fig:cumu_return}
\end{figure}

\begin{figure}[H]
    \centering
    \includegraphics[width=0.8\linewidth]{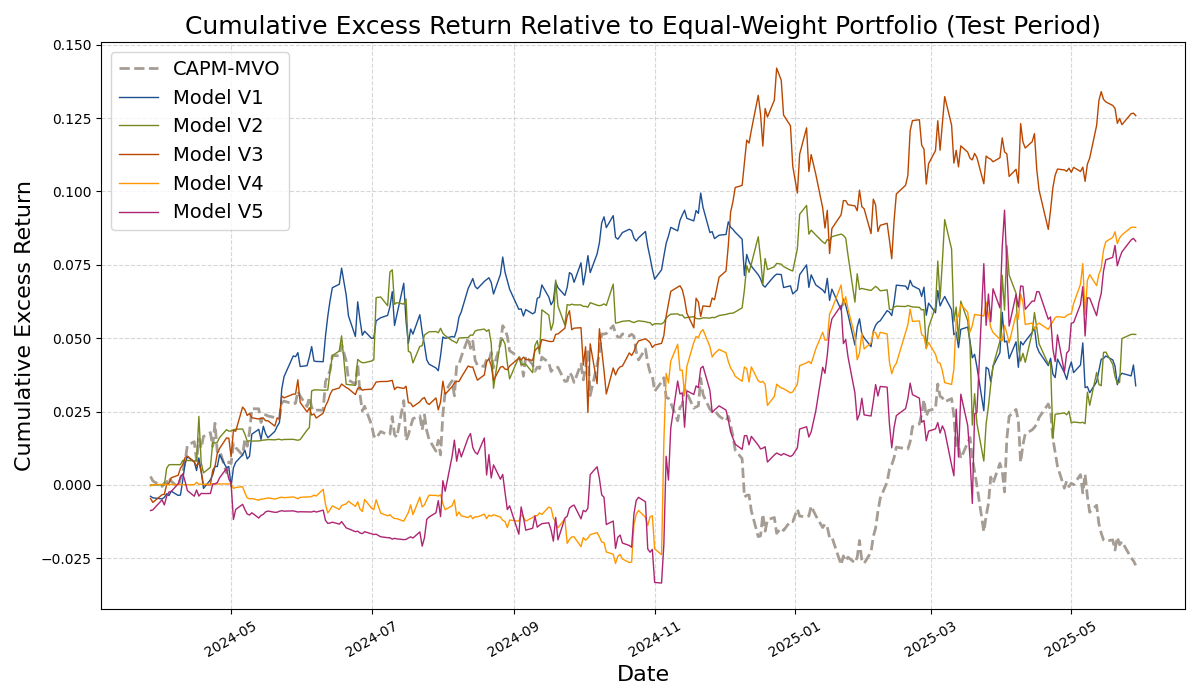}
    \caption{Cumulative excess return relative to the equal-weight portfolio (test period).}
    \label{fig:excess_return}
\end{figure}

Figures~\ref{fig:cumu_return} and~\ref{fig:excess_return} illustrate the performance of our proposed models compared to the equal-weight and CAPM-MVO baselines during the test period. Figure~\ref{fig:cumu_return} presents the cumulative return over time of all portfolios, and Figure~\ref{fig:excess_return} shows the cumulative excess returns relative to the equal-weight benchmark.The evaluation metrics for all models are reported in Table~\ref{metrics}.

\begin{table}[H]

\centering
\small
\caption{Performance Comparison (Test Period)}
\label{metrics}
\resizebox{\linewidth}{!}{%
\begin{tabular}{lrrrrrrr}

\toprule
\textbf{Metric} & \textbf{Model v1} & \textbf{Model v2} & \textbf{Model v3} & \textbf{Model v4} & \textbf{Model v5} & \textbf{Equal-Weight} & \textbf{CAPM-MVO} \\
\midrule
Total Return         & 28.11\% & 29.86\% & \textcolor{DodgerBlue}{\textbf{37.32\%}} & 33.50\% & 33.04\% & 24.73\% & 21.99\% \\
Annualized Return    & 23.65\% & 25.10\% & \textcolor{DodgerBlue}{\textbf{31.23\%}} & 28.10\% & 27.72\% & 20.85\% & 18.58\% \\
Volatility           & \textcolor{DodgerBlue}{\textbf{26.02\%}} & 26.29\% & 27.19\% & 26.60\% & 28.45\% & 24.89\% & 22.03\% \\
Sharpe Ratio         & 0.91    & 0.95    & \textcolor{DodgerBlue}{\textbf{1.15}}    & 1.06    & 0.98    & 0.83    & 0.84    \\
VaR (95\%)           & -2.62\% & -2.57\% & \textcolor{DodgerBlue}{\textbf{-2.5\%}} & -2.68\% & -2.64\% & -2.53\% & -2.04\% \\
Max Drawdown         & -23.38\% & -23.35\% & -22.05\% & -21.70\% & \textcolor{DodgerBlue}{\textbf{-20.99\%}} & -22.60\% & -21.59\% \\
\bottomrule
\end{tabular}
}

\vspace{1mm}
\begin{minipage}{0.95\textwidth}
\footnotesize\textit{Note: Values highlighted in blue represent the best performance across all models for each respective metric.}
\end{minipage}
\end{table}
Table~\ref{metrics} reports the performance of five model variants, along with our benchmarks. Across the board, all LSTM-GAT models outperform the benchmarks, demonstrating the effectiveness of our end-to-end LSTM-GAT framework. A detailed percentage comparison against the benchmarks is provided in Appendix~\ref{tab:model_comparsion}.

\textbf{Model v1} with only three price-based features, already delivers a 13.43\% higher annualized return and a 9.64\% higher Sharpe ratio than the equal-weight portfolio, and a 27.28\% higher annualized return with an 8.3\% higher Sharpe ratio relative to the CAPM-MVO benchmark. It also has the lowest volatility of 26.02\% among all model variants, indicating relative stability. However, its limited features result in the weakest return performance among the 5 model variants.

\textbf{Model v2} demonstrates notable improvements by incorporating additional price-based features, improving annualized return and Sharpe ratio by 20.38\% and 14.46\% over equal-weight, and by 35.1\% and 13.1\% over CAPM-MVO, respectively. However, this comes with increased volatility, suggesting that more price features enhance signal capture but may also introduce noise.

 \textbf{Model v3}  introduces sentiment-based features and achieves the best overall performance: a 31.23\% annualized return and 1.15 Sharpe ratio, representing improvements of nearly 50\% and 39\% over equal-weight, and over 68\% and 37\% over CAPM-MVO. It also records the lowest Value at Risk (VaR) at -2.5\%, demonstrating strong downside protection. These gains validate the value of integrating financial sentiment signals.
 
\textbf{Model v4} replaces the static graph with a dynamic one, resulting in 2.1\% lower volatility and improved drawdown, but it slightly underperforms v3 in returns. This suggests that dynamic graphs help capture evolving relationships, but may capture short-lived or spurious relationships.

 \textbf{Model v5} applies PCA for dimensionality reduction and achieves the best performance in terms of drawdown control, with a maximum drawdown of only -20.99\% during the April 2025 tariff shock. Although its annualized return and Sharpe ratio are slightly lower than those of \textbf{Model v3} and \textbf{Model v4}, this result suggests PCA can filter out noise and improve portfolio resilience under adverse market conditions.
 
Our analysis demonstrates that expanding the feature set and incorporating financial news sentiment consistently improves model performance. \textbf{Model v3}, which includes both price-based and sentiment features, achieves the best overall returns and risk-adjusted metrics. While \textbf{Model v4} introduces a dynamic graph structure that reduces volatility, and \textbf{Model v5} leverages PCA to further reduce max drawdown, both trade off some return for improved risk control. These results highlight the importance of combining diverse signals and advanced graph structures to build robust, high-performing portfolio models.

\subsection{Discussion and Significance}

Table~\ref{tab:model_comparison} compares the performance of our end-to-end LSTM-GAT framework with two representative deep learning approaches from the literature (see Table~\ref{tab:model_comp} in Appendix~\ref{comp} for details of model design).

\begin{table}[H]
\centering
\caption{Comparison of Model Design and Performance}
\begin{tabular}{L{3.3cm} L{3.6cm} L{3.6cm} L{3.6cm}}
\toprule
\textbf{} & \textbf{Our Model} & \textbf{Chalvatzis (2019)\cite{chalvatzis2019high}} & \textbf{Lu et al. (2025)~\cite{lu2025leveraging}} \\
\midrule
\textbf{Time period }
& Jan 2021 – May 2025
& Jan 2010 – Apr 2018
& Aug 2023 – Dec 2023 \\

\textbf{Annualized return }
& 31.23\%
& 19.50\%
& 302.47\% \\

\textbf{Sharpe ratio (annualized)}
& 1.15
& 0.28
& 0.85 \\

\textbf{Max drawdown}
& -20.99\%
& -20.00\%
& -3.79\% \\
\bottomrule
\end{tabular}
\label{tab:model_comparison}
\end{table}

Compared to earlier LSTM-based portfolio allocation models~\cite{chalvatzis2019high}, our framework achieves stronger results both in annualized return and Sharpe ratio. Their best-performing model reported an annualized return of 20.3\% and a Sharpe ratio of 0.28, while our approach achieves higher values across both metrics. We attribute these gains primarily to two enhancements: (i) the use of a Graph Attention Network (GAT), which allows the model to incorporate inter-stock dependencies, and (ii) the addition of sentiment features, which bring in information on market psychology often missing from purely price-based strategies. Together, these components enable more informed and adaptive allocation decisions.

Lu et al.\cite{lu2025leveraging} propose a two-step framework combining a bidirectional LSTM and graph structure. Their reported annualized returns are extremely high (160–300\%), but the corresponding Sharpe ratios are relatively modest at around 0.85. These conditions may greatly inflate simulated returns. In addition, these results may reflect the very short and unusually favorable evaluation window (August-December 2023), as well as possible overfitting to a small sample. In addition, their method first predicts the next day’s closing prices for all stocks and then adjust their portfolio based on these predictions, which may lead to unstable portfolio decisions and extreme results. Lu et al. mentions that their results are based solely on this selected historical backtests; thus their models' performance are not verified  under more realistic or more unpredictable market periods. By contrast, our evaluation covers more than a year and includes a major stress event (the April 2025 tariff shock). While this naturally results in lower raw returns, our models achieve higher Sharpe ratios (0.91-1.15), indicating stronger risk-adjusted performance and robustness in volatile conditions. Notably, even our price-only models (v1 and~v2) outperform Lu et al.’s best model in terms of Sharpe ratio, despite using similar or fewer features.

Beyond numerical comparisons, the results provide insight into economic mechanisms. Incorporating sentiment signals proved especially valuable: Model v3, which includes sentiment, delivered the highest returns and Sharpe ratio. This suggests that market mood, as captured in financial news, provides predictive information not fully embedded in prices. Similarly, Model v5, which applies PCA to reduce dimensionality, produced the lowest maximum drawdown during the 2025 shock. This highlights PCA’s role in filtering out noisy or redundant features, making allocations more stable under stress.

From a practitioner’s perspective, the framework demonstrates how modern deep learning methods can be adapted for portfolio management. A manager could, for example, use sentiment-augmented models to complement traditional signals, especially around event-driven volatility. PCA-based dimensionality reduction could be employed when robustness is prioritized over maximizing returns. Importantly, by optimizing directly for Sharpe ratio, the model aligns its objective with the performance criteria most relevant to investors.

These findings underscore the advantage of end-to-end portfolio learning. By integrating temporal dynamics, cross-asset dependencies, and sentiment in a single framework, our approach avoids the instability of two-step prediction-optimization pipelines and produces more resilient performance across diverse market conditions.

\section{Limitations and Improvements}
The current model has several limitations that we will address in future work.

First, we limited the model to a fixed portfolio of nine stocks from the S\&P~500. While this simplifies the experiment setup, it constrains GAT, which works best when the graph has enough nodes to capture rich relationships; with only nine stocks, the graph may be too sparse to learn useful patterns. A larger and more diverse equity universe would yield a denser, more informative graph and enable more expressive relational learning. As a next step, we plan to expand the stock universe by including firms outside the S\&P~500 and conducting more comprehensive correlation tests to construct a more diversified portfolio universe, leading to richer graph relationships and improved generalizability. For small-cap firms with limited financial news coverage, we may also consider extracting sentiment from social media sources such as X and Reddit.

Additionally, we obtained sentiment scores from third-party APIs whose methods, data sources, and training processes are not clearly documented. This lack of transparency may reduce the accuracy of our sentiment features. Therefore, we will develop a custom sentiment pipeline, such as fine-tuning FinBERT or using a large language model (LLM) to extract sentiment from financial news, to increase control and interpretability.

Moreover, the current model makes several simplifying assumptions. It assumes ideal trading conditions by ignoring transaction costs, liquidity constraints, and execution delays. While common in early-stage research, this can produce overly optimistic results, especially in the context of large-volume institutional trading. To make the evaluation more realistic and scalable, we will include trading costs and slippage. 

Furthermore, we ran experiments on standard CPU-based machines with limited computing power, which restricted hyperparameter exploration and the training of more complex models. As a result, the reported performance may underestimate the method’s potential. Using more capable hardware—GPUs or cloud-based clusters—will allow faster experiments and better tuning.

The model also assumes static sector classifications based on GICS over the five-year period, even though companies can change sectors due to mergers, acquisitions, or strategic shifts. This static assumption may introduce structural inaccuracies in the graph. We will incorporate time-varying sector data or design adaptive graph-construction mechanisms to address this.

Lastly, our dynamic-graph construction relies on a fixed correlation threshold of 0.5 with binary edge weights. Although computationally convenient, this heuristic may discard information in the magnitude of correlations. We will explore treating the threshold as a tunable hyperparameter, using weighted edges, and adopting alternative similarity metrics (e.g., mutual information) to build more robust graph structures.

\section{Conclusion}
This study explored an end-to-end deep learning framework for portfolio optimization that integrates three key elements: temporal modeling with LSTMs, relational modeling with GATs, and sentiment analysis from financial news. By combining these components in a unified pipeline, the model generates portfolio weights directly, avoiding compounded errors common in two-step prediction-optimization approaches.

Our findings suggest that incorporating both sentiment and graph-based dependencies leads to more robust, risk-adjusted performance, even under volatile conditions such as the 2025 tariff shock. Sentiment features improved returns and Sharpe ratios, while dimensionality reduction via PCA helped stabilize allocations and limit drawdowns. Together, these results highlight the potential of modern deep learning methods to complement or enhance traditional portfolio strategies.

At the same time, the study is exploratory in scale, relying on a fixed nine-stock universe and simplified assumptions such as zero transaction costs. These limitations make the results best viewed as a proof of concept. Future work will extend the approach to larger and more diverse asset universes, incorporate market frictions and liquidity effects, and develop more transparent sentiment pipelines.

In summary, the proposed framework demonstrates that blending deep learning with alternative data can produce more adaptive and resilient portfolio strategies, opening new directions for quantitative asset management.

\newpage
\printbibliography

\newpage
\section{Appendix}

\subsection{Feature Description}\label{feature}

\begin{table}[H]
\centering
\caption{Price-Based Features}
\begin{tabular}{p{3.2cm} p{7cm} p{4.5cm}}
\toprule
\textbf{Feature} & \textbf{Description} & \textbf{Calculation} \\
\midrule
Open & Stock price at the start of the trading day & Raw data from API \\
High & Highest price during the trading day & Raw data from API \\
Low & Lowest price during the trading day & Raw data from API \\
Close & Stock price at market close & Raw data from API \\
Volume & Total number of shares traded in the day & Raw data from API \\
Log Return & Logarithmic change in closing price from day $t{-}1$ to $t$ & $\log\left(\frac{P_t}{P_{t-1}}\right)$ \\
Annualized Returns (1W/2W/1M) & Returns over 1-week, 2-week, or 1-month periods, scaled to yearly rate & $r_{\text{period}} \times \frac{252}{\text{days}}$ \\
1W Rolling Volatility & Standard deviation of daily log returns over a 5-day window & $\text{std}(\text{Log Return}_{t-4:t})$ \\
MACD (1W–1M) &
Momentum indicator calculated as the difference between short-term (1W) and long-term (1M) EMAs &
\begin{tabular}[t]{@{}l@{}}
$\text{MACD}_t = \text{EMA}_{5,t} - \text{EMA}_{21,t}$ \\
$\text{EMA}_t = \alpha P_t + (1{-}\alpha)\text{EMA}_{t-1}$ \\
$\alpha = \frac{2}{N + 1}$
\end{tabular} \\
\bottomrule
\end{tabular}
\end{table}

\begin{table}[H]
\centering
\caption{Sentiment-Based Features}
\begin{tabular}{p{3.2cm} p{7cm} p{4.5cm}}
\toprule
\textbf{Feature} & \textbf{Description} & \textbf{Calculation} \\
\midrule
News Count & Total number of articles related to the stock on a given day & Count of matched news entries \\
News Frequency & Proportion of news about a stock relative to all stocks that day & $\frac{\text{Stock News Count}}{\text{Total News Count}}$ \\
Average Sentiment & Mean sentiment score across all articles for the day & $\frac{1}{N} \sum_{i=1}^{N} s_i$ \\
Sentiment Variance & Variability of sentiment scores across articles for the day & $\text{Var}(s_1, s_2, ..., s_n)$ \\
Weighted Sentiment & Adjusted sentiment based on news frequency and average sentiment & $\text{News Freq.} \times \text{Avg. Sentiment}$ \\
\bottomrule
\end{tabular}
\end{table}

\subsection{Hyperparameters}
\begin{table}[H]
\centering
\caption{Explanation of Hyperparameters Used in LSTM-GAT Portfolio Model}
\begin{tabular}{p{3.5cm} p{10.7cm}}
\toprule
\textbf{Hyperparameter} & \textbf{Explanation} \\
\midrule
\texttt{batch\_size} & Number of samples used in each training iteration. Smaller sizes can generalize better but are noisier. \\
\texttt{lstm\_hidden} & Number of hidden units in the LSTM layer. Controls how much information the LSTM retains. \\
\texttt{lstm\_layers} & Number of stacked LSTM layers. More layers can capture more complex time dependencies. \\
\texttt{lstm\_dropout} & Dropout rate applied to the LSTM layer to reduce overfitting. \\
\texttt{lstm\_bidirectional} & Boolean flag for whether the LSTM is bidirectional (processes input forward and backward). \\
\texttt{gat\_hidden} & Number of hidden units in the Graph Attention Network (GAT) layer. \\
\texttt{gat\_dropout} & Dropout rate applied to the GAT layer for regularization. \\
\texttt{gat\_alpha} & Negative slope coefficient for the LeakyReLU activation in the GAT layer. \\
\texttt{final\_dropout} & Dropout rate applied before the final prediction layer. \\
\texttt{learning\_rate} & Controls how quickly the model updates during training. Smaller values lead to slower but more stable training. \\
\texttt{lstm\_weight\_decay} & L2 regularization strength for the LSTM layer to prevent overfitting. \\
\texttt{gat\_weight\_decay} & L2 regularization strength for the GAT layer. \\
\texttt{final\_weight\_decay} & L2 regularization strength for the final dense layer. \\
\bottomrule
\end{tabular}
\end{table}

\begin{table}[h!]
\centering
\caption{Hyperparameter search space used in Optuna trials}
\label{tab:hyperparam_ranges}
\begin{tabular}{p{4cm} p{3.5cm} p{4cm} p{2cm}}
\toprule
\textbf{Hyperparameter} & \textbf{Type} & \textbf{Range / Values} & \textbf{Step} \\
\midrule
\texttt{batch\_size}  & Categorical & \{16, 32, 64\} & -- \\
\texttt{lstm\_hidden} & Integer & 32 -- 128 & 16 \\
\texttt{lstm\_layers} & Integer & 1 -- 3 & 1 \\
\texttt{lstm\_dropout} & Float & 0.0 -- 0.5 & 0.05 \\
\texttt{lstm\_bidirectional} & Fixed & False & -- \\
\texttt{gat\_hidden} & Integer & 32 -- 128 & 16 \\
\texttt{gat\_layers} & Fixed & 2 & -- \\
\texttt{gat\_heads} & Fixed & 1 & -- \\
\texttt{gat\_dropout} & Float & 0.1 -- 0.5 & 0.05 \\
\texttt{gat\_alpha} & Float & 0.05 -- 0.3 & 0.05 \\
\texttt{final\_dropout} & Float & 0.1 -- 0.5 & 0.05 \\
\texttt{learning\_rate} & Log-uniform float & $10^{-4}$ -- $10^{-2}$ & log \\
\texttt{lstm\_weight\_decay} & Log-uniform float & $10^{-6}$ -- $10^{-2}$ & log \\
\texttt{gat\_weight\_decay} & Log-uniform float & $10^{-6}$ -- $10^{-2}$ & log \\
\texttt{final\_weight\_decay} & Log-uniform float & $10^{-6}$ -- $10^{-2}$ & log \\
\bottomrule
\end{tabular}
\end{table}

\begin{table}[H]
\centering
\caption{Hyperparameters Used in Models}
\small
\begin{tabular}{
    p{3.8cm} 
    >{\raggedleft\arraybackslash}p{1.68cm} 
    >{\raggedleft\arraybackslash}p{1.68cm} 
    >{\raggedleft\arraybackslash}p{1.68cm} 
    >{\raggedleft\arraybackslash}p{1.68cm} 
    >{\raggedleft\arraybackslash}p{1.68cm}
}
\toprule
\textbf{hyperparameter} & \textbf{Model V1} & \textbf{Model V2} & \textbf{Model V3} & \textbf{Model V4} & \textbf{Model V5} \\
\midrule
\texttt{batch\_size} & 64 & 32 & 32 & 64 & 32 \\
\texttt{lstm\_hidden} & 96 & 96 & 32 & 80 & 32 \\
\texttt{lstm\_layers} & 2 & 3 & 2 & 1 & 1 \\
\texttt{lstm\_dropout} & 0.10 & 0.25 & 0.0 & 0.27 & 0.21 \\
\texttt{gat\_hidden} & 96 & 64 & 64 & 80 & 32 \\
\texttt{gat\_dropout} & 0.10 & 0.25 & 0.30 & 0.20 & 0.25 \\
\texttt{gat\_alpha} & 0.10 & 0.30 & 0.25 & 0.15 & 0.35 \\
\texttt{lstm\_weight\_decay} & 5.71e-04 & 9.44e-05 & 1.08e-06 & 3.33e-03 & 1.99e-04 \\
\texttt{gat\_weight\_decay} & 1.68e-05 & 1.23e-04 & 2.78e-03 & 2.48e-04 & 5.54e-04 \\
\texttt{learning\_rate} & 7.02e-04 & 3.48e-03 & 1.27e-03 & 3.98e-03 & 1.41e-03 \\
\texttt{final\_dropout} & 0.20 & 0.35 & 0.25 & 0.29 & 0.34 \\
\texttt{final\_weight\_decay} & 1.74e-04 & 5.13e-05 & 2.00e-03 & 2.69e-04 & 5.00e-04 \\
\texttt{lstm\_bidirectional} & False & False & False & False & False \\
\bottomrule
\end{tabular}
\vspace{1mm}

\footnotesize{\textit{Note: Decimal values have been rounded for readability.}}
\end{table}

\subsection{Libraries Used}
\begin{table}[H]
\centering
\caption{Libraries Used in the Project}
\begin{tabular}{p{3.5cm} p{10.7cm}} 
\toprule
\textbf{Library} & \textbf{Description} \\
\midrule
\texttt{matplotlib} & Used to visualize model performance, including cumulative returns and weight paths. \\
\texttt{NumPy} & Performs numerical operations like matrix calculations, statistics, and return processing. \\
\texttt{os} & Handles file system operations such as reading data and setting environment variables. \\
\texttt{pandas} & Loads, cleans, merges, and manages tabular time-series data for stocks and news. \\
\texttt{PyTorch} & Core deep learning framework used to implement LSTM and GAT models. \\
\texttt{random} & Ensures reproducibility by controlling random shuffling and seed setting. \\
\texttt{scikit-learn} & Used for feature standardization via \texttt{StandardScaler} to prepare input data for modeling. \\
\texttt{tqdm} & Adds progress bars to loops during training and testing for better tracking. \\
\bottomrule
\end{tabular}
\end{table}

\subsection{CAPM-MVO Baseline Strategy} 

To benchmark the performance of our proposed LSTM-GAT framework, we implemented a classical \textbf{Capital Asset Pricing Model (CAPM)-based Mean-Variance Optimization (MVO)} strategy. This approach estimates asset returns using CAPM and determines portfolio weights by maximizing the Sharpe ratio under modern portfolio theory.

\subsubsection*{CAPM-Based Return Estimation} \label{CAPM}

Expected returns were calculated using rolling 252-day windows and a 21-day rebalance frequency. For each window, we estimated stock betas relative to market excess returns via linear regression. The CAPM expected return for each asset $i$ was computed as:
\begin{equation}
    \mathbb{E}[R_i] = R_f + \beta_i (\mathbb{E}[R_m] - R_f)
\end{equation}

where $R_f$ is the annualized risk-free rate, and $\mathbb{E}[R_m]$ is the expected market return. When CAPM-predicted returns underperformed the risk-free rate for all assets in a window, we reverted to historical mean returns as a fallback.

\subsubsection*{Portfolio Construction}

Portfolio weights were optimized using the \textbf{maximum Sharpe ratio} objective:
\begin{equation}
    \max_w \frac{w^\top \mu - R_f}{\sqrt{w^\top \Sigma w}}
\end{equation}

subject to:
\begin{itemize}
    \item $\sum_i w_i = 1$ (fully invested)
    \item $w_i \in [-1.5, 1.5]$ (long/short allowed)
\end{itemize}

If this optimization failed, we defaulted to the Global Minimum Variance (GMV) portfolio for that period. Weights were recalculated every 21 trading days and held constant between rebalancing points.

\subsection{Model Comparison with Benchmark}
\label{tab:model_comparsion}
\begin{table}[h!]
\centering
\caption{Percentage Difference of Models v1--v5 Compared to Equal-Weight}
\label{tab:equalweight-comparison}
\begin{tabular}{p{3.8cm} 
    >{\raggedleft\arraybackslash}p{1.68cm} 
    >{\raggedleft\arraybackslash}p{1.68cm} 
    >{\raggedleft\arraybackslash}p{1.68cm} 
    >{\raggedleft\arraybackslash}p{1.68cm} 
    >{\raggedleft\arraybackslash}p{1.68cm}}
\toprule
\textbf{Metric} & \textbf{Model v1} & \textbf{Model v2} & \textbf{Model v3} & \textbf{Model v4} & \textbf{Model v5} \\
\midrule
Total Return & 13.67\% & 20.74\% & 50.91\% & 35.46\% & 33.60\% \\
Annualized Return & 13.43\% & 20.38\% & 49.78\% & 34.77\% & 32.95\% \\
Volatility & 4.54\% & 5.62\% &  9.24\% & 6.87\% & 14.30\% \\
Sharpe Ratio & 9.64\% & 14.46\% & 38.55\% & 27.71\% & 18.07\% \\
VaR (95\%) & 3.56\% & 1.58\% & -1.19\% &  5.93\% &  4.35\% \\
Max Drawdown & 3.45\% & 3.32\% & -2.43\% & -3.98\% & -7.12\% \\
\bottomrule
\end{tabular}
\end{table}

\begin{table}[h!]
\centering
\caption{Percentage Difference of Models v1--v5 Compared to CAPM-MVO}
\label{tab:capm-comparison}
\begin{tabular}{p{3.8cm} 
    >{\raggedleft\arraybackslash}p{1.68cm} 
    >{\raggedleft\arraybackslash}p{1.68cm} 
    >{\raggedleft\arraybackslash}p{1.68cm} 
    >{\raggedleft\arraybackslash}p{1.68cm} 
    >{\raggedleft\arraybackslash}p{1.68cm}}
\toprule
\textbf{Metric} & \textbf{Model v1} & \textbf{Model v2} & \textbf{Model v3} & \textbf{Model v4} & \textbf{Model v5} \\
\midrule
Total Return & 27.83\% & 35.79\% & 69.71\% & 52.34\% & 50.25\% \\
Annualized Return & 27.29\% & 25.09\% & 68.08\% & 51.24\% & 49.19\% \\
Volatility & 18.11\% & 19.34\% &  23.42\% & 20.74\% & 29.14\% \\
Sharpe Ratio & 8.33\% & 13.10\% & 36.90\% & 26.19\% & 16.67\% \\
VaR (95\%) & 28.43\% & 25.98\% & 22.55\% &  31.37\% &  29.41\% \\
Max Drawdown & 8.29\% & 8.15\% & 2.13\% & 0.51\% & -2.78\% \\
\bottomrule
\end{tabular}
\end{table}

\subsection{Comparison with Existing Literature} \label{comp}
\begin{table}[H]
\centering
\caption{Comparison of Model Design and Performance}
\begin{tabular}{L{3.3cm} L{3.6cm} L{3.6cm} L{3.6cm}}
\toprule
\textbf{Aspect} & \textbf{Our Model} & \textbf{Chalvatzis \& Hristu-Varsakelis (2019)} & \textbf{Lu et al. (2025)} \\
\midrule

\textbf{Model design }
& LSTM-GAT with sentiment-based dynamic graph; direct portfolio weights
& LSTM model for stock index trend prediction
& BiLSTM-GAT-AM using dual graphs and attention mechanism; two-step framework \\

\textbf{Stock universe}
& Nine fixed stocks selected from the S\&P~500
& S\&P~500 ETF
& Top N stocks from 82 S\&P~500 by predicted next-day return \\

\textbf{Input features}
& Price-based \& sentiment-based features
& Price-based features
& Price-based features \\

\textbf{Optimization Function}
& Maximize Sharpe ratio
& Maximize predicted profit
& Maximize Sharpe ratio\\

\textbf{Time period }
& Jan 2021 – May 2025
& Jan 2010 – Apr 2018
& Aug 2023 – Dec 2023 \\

\textbf{Annualized return }
& 31.23\%
& 19.50\%
& 302.47\% \\

\textbf{Sharpe ratio}
& 1.15
& 0.28
& 0.85 \\

\textbf{Max drawdown}
& -20.99\%
& -20.00\%
& -3.79\% \\
\bottomrule
\end{tabular}
\label{tab:model_comp}
\end{table}

\subsection{Model Predicted Weights}

\begin{figure}[H]
    \centering
    \includegraphics[width=1\linewidth]{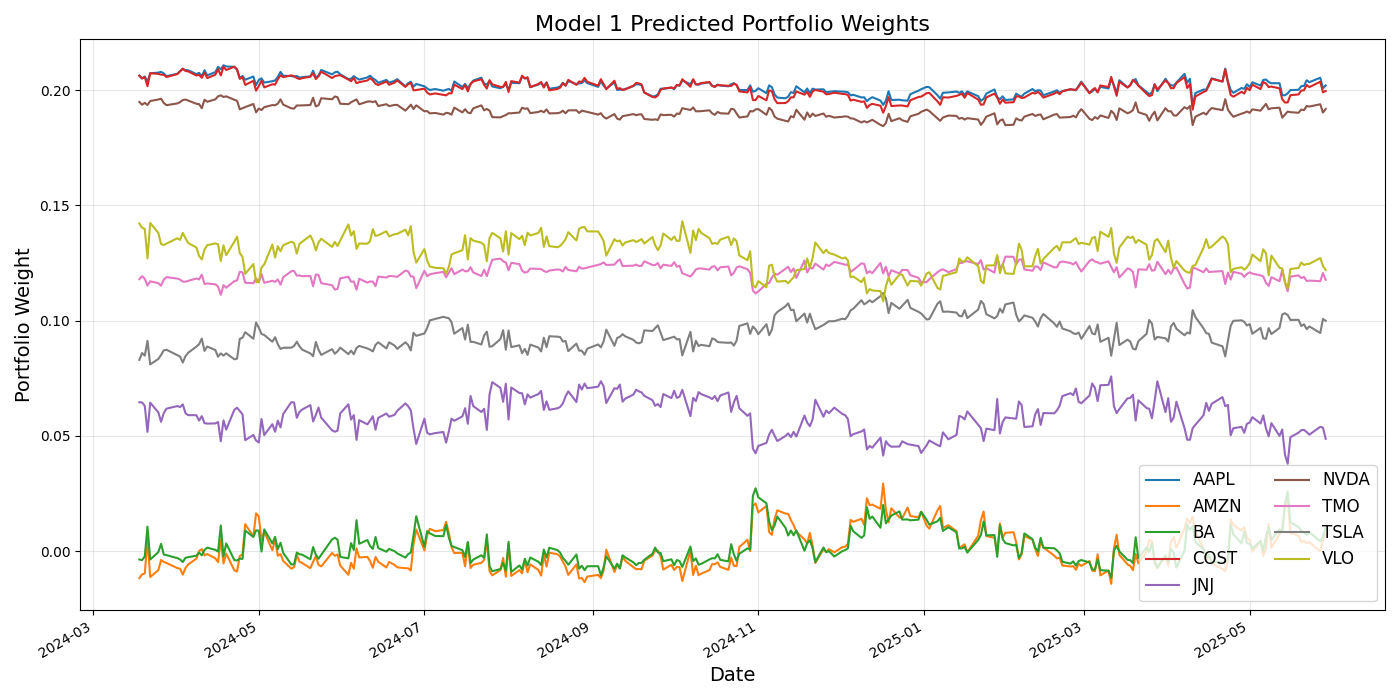}
    \caption{Model v1}
    \label{fig:predicted_weights_M1}
\end{figure}
\begin{figure}
    \centering
    \includegraphics[width=1\linewidth]{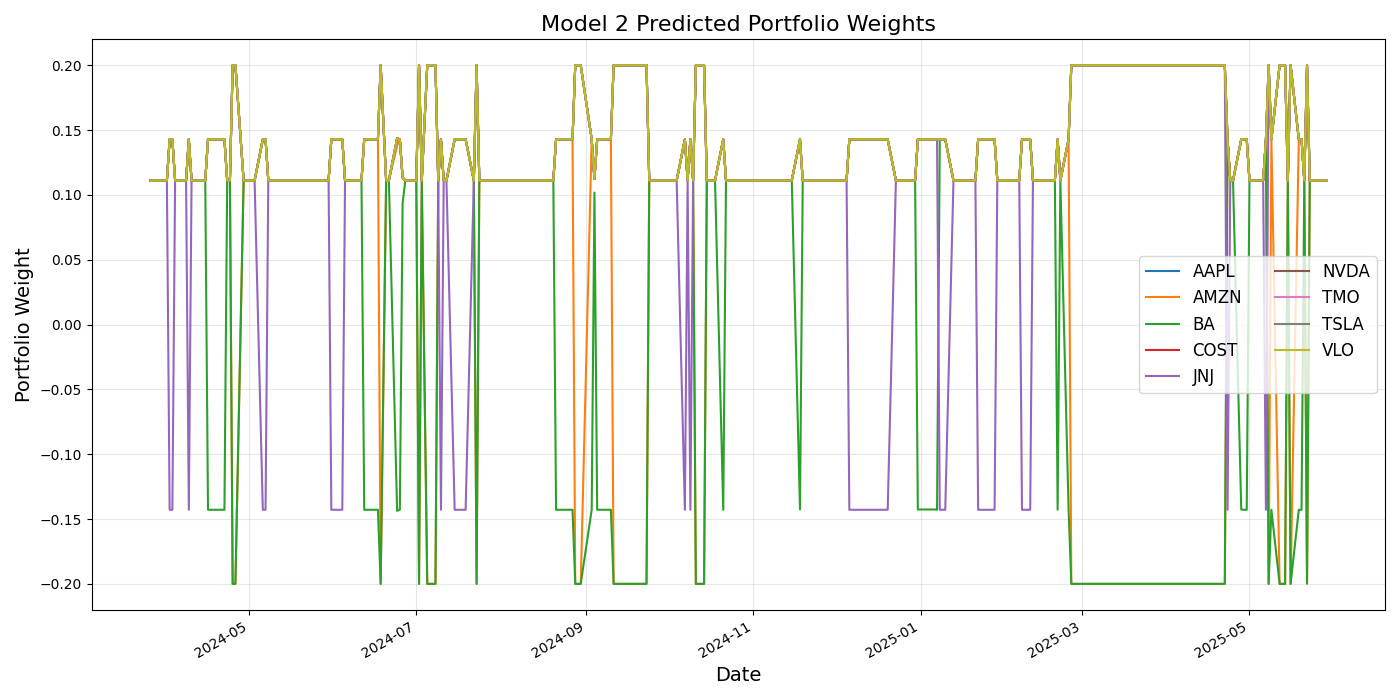}
    \caption{Model v2}
    \label{fig:predicted_weights_M2}
\end{figure}
\begin{figure}
    \centering
    \includegraphics[width=1\linewidth]{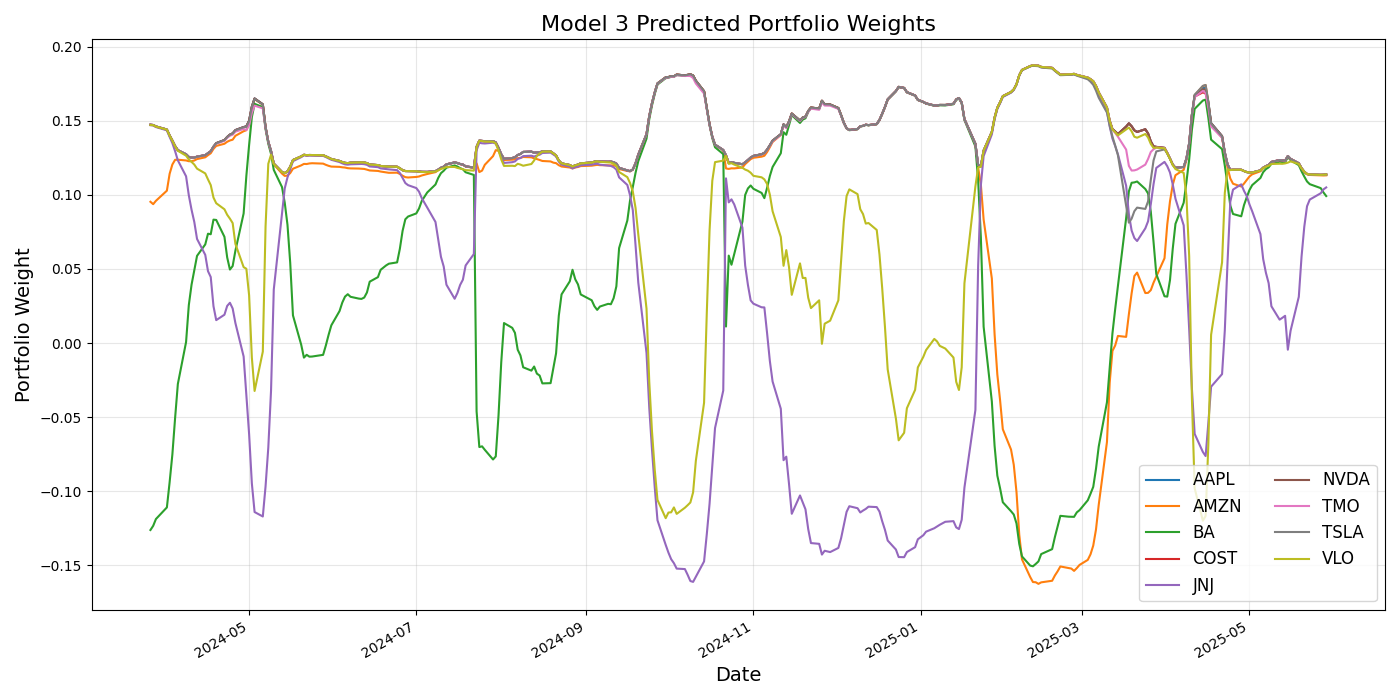}
    \caption{Model v3}
    \label{fig:predicted_weights_M3}
\end{figure}
\begin{figure}
    \centering
    \includegraphics[width=1\linewidth]{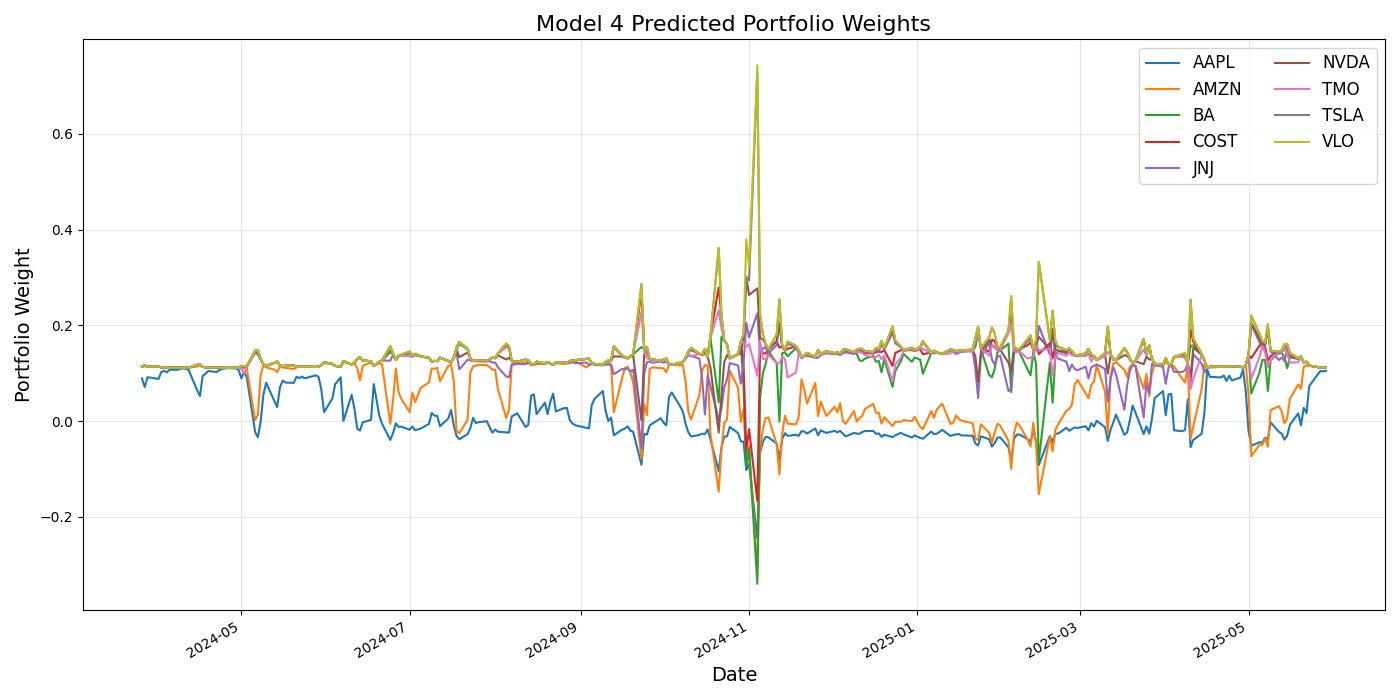}
    \caption{Model v4}
    \label{fig:predicted_weights_M4}
\end{figure}
\begin{figure}[H]
    \centering
    \includegraphics[width=1\linewidth]{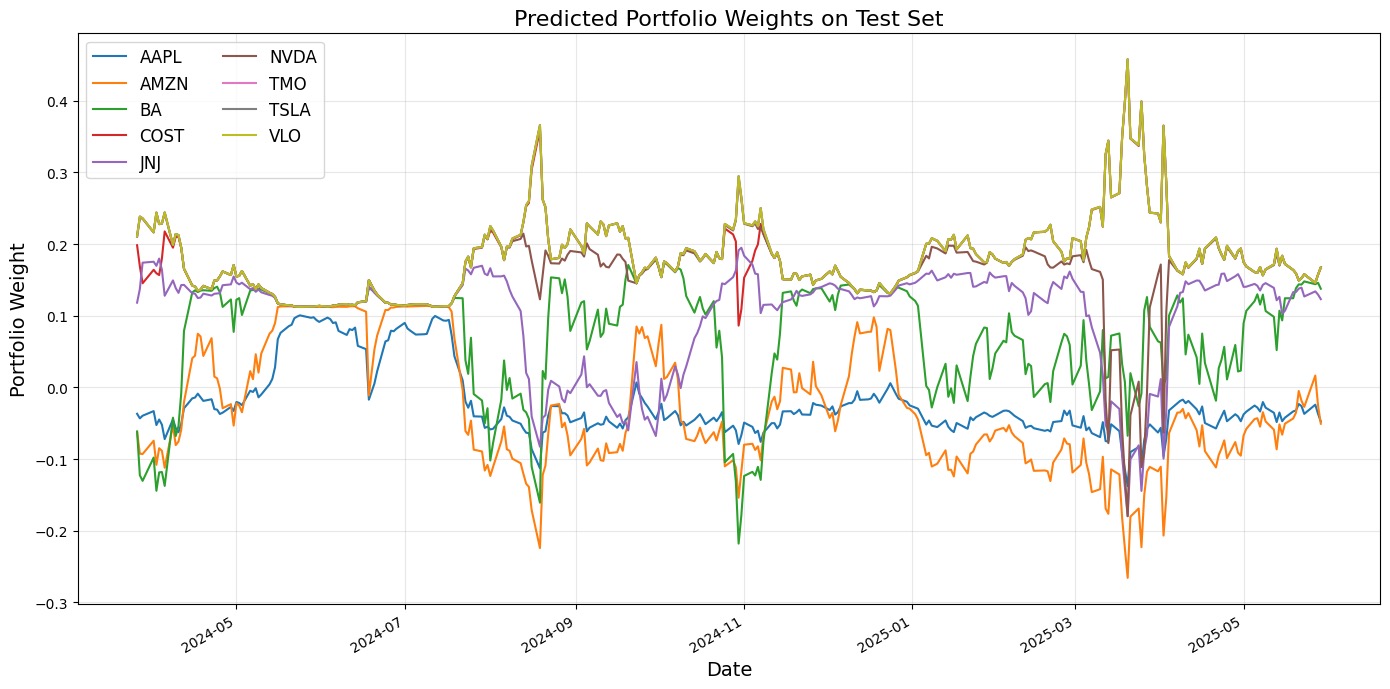}
    \caption{Model v5}
    \label{fig:predicted_weights_M5}
\end{figure}

\end{document}